\documentclass[aps,prb,twocolumn,floatfix,superscriptaddress,
showkeys,amsmath,amssymb,longbibliography]{revtex4-2}		
 
\usepackage{braket}
\usepackage{graphicx}
\usepackage{dcolumn}
\usepackage{bm}
\usepackage[colorlinks,urlcolor=blue,citecolor=blue,linkcolor=blue]{hyperref}
\usepackage{subfig}
\usepackage[font=small,labelfont=bf,format=plain,justification=centerlast,labelsep=period]{caption}
\usepackage{tikz}
\usepackage{comment}
\usetikzlibrary{arrows,shapes,positioning,shadows,svg.path,backgrounds,fit}

\usepackage{nicefrac}
\usepackage{verbatim}
\usepackage[normalem]{ulem}
\usepackage{orcidlink}

\newcommand{\gammahat}{\hat{\gamma}}
\newcommand{\ahat}{\hat{a}}
\newcommand{\Xhat}{\hat{X}}
\newcommand{\Yhat}{\hat{Y}}
\newcommand{\rr}{\mathbf{r}}
\newcommand{\kk}{\mathbf{k}}
\newcommand{\qq}{\mathbf{q}}

\begin{document}


\title{Moiré-Polaritons in a Dark Bose-Einstein Condensate}

\author{Moroni Santiago-Garc\'ia \orcidlink{0000-0002-6321-4853
}}
\affiliation{
Instituto Nacional de Astrof\'isica, \'Optica y Electr\'onica, Calle Luis Enrique Erro 1, Sta. Ma.  Tonantzintla, Puebla CP 72840, M\'exico}

\author{Shunashi G.\ Castillo-L\'opez
\orcidlink{0000-0002-2251-3961}}
\author{David A.\ Ruiz-Tijerina \orcidlink{0000-0001-7688-2511}}
\author{Arturo Camacho-Guardian
\orcidlink{0000-0001-5161-5468
}}

\email{acamacho@fisica.unam.mx}
\affiliation{Instituto de F\'isica, Universidad Nacional Aut\'onoma de M\'exico, Apartado Postal 20-364, Ciudad de M\'exico C.P. 01000, M\'exico}
\date{\today}

\begin{abstract}
Quantum mixtures of moiré excitons have arisen as a platform for realizing novel phases of light and matter. Here, we study moiré polaritons coupled to a Bose-Einstein condensate of dark-state excitons confined to a moiré superlattice. We develop a variational approach to analyze the optical response of the system and demonstrate that strong exciton-exciton interactions significantly modify the character of moiré polaritons, leading to sizable energy shifts of the avoided crossing between the principal polariton branches, and the emergence of an additional, stable repulsive-polariton bound state.

\end{abstract}

\maketitle

{\it Introduction.-} Moiré heterostructures have emerged as a powerful platform for engineering strongly correlated phases of light and matter with unprecedented tunability~\cite{Andrei_2020}. Twistronic control over the interplay between kinetic energy and interactions has enabled the experimental realization of intriguing electronic phases, including unconventional superconductivity~\cite{Cao_2018_unc_sc,Chen_2019_sc}, Mott insulators~\cite{Cao_2018,Chen_2019}, Wigner crystals~\cite{Padhi_2018,Padhi_2019,Regan_2020}, and topological states~\cite{Spanton_2018}. This has positioned moiré structures as a versatile and scalable architecture for quantum simulation and the exploration of novel electronic phenomena~\cite{Balents_2020,Wang_2020,Tang_2020,Kennes_2021}.

Engineering the electronic bands with moiré superlattices also modifies their optical response, unveiling new classes of moiré excitons and moiré exciton-polaritons with no counterpart in conventional semiconductors~\cite{Dufferwiel_2015,Yu_2017,Jin_2019,Huang_2022}. The confinement of excitons to an emergent moir\'e superlattice can enhance their lifetime~\cite{Miao_2021}, induce strong exciton-exciton interactions~\cite{Datta_2022,Kumlin_2025}, and promote novel collective quantum phases, such as dark Bose-Einstein condensates~\cite{Lagoin_2021,Cooper_2022}, dipolar ladders~\cite{Park_2023}, moiré-induced nonlinearities~\cite{Zhang_2021,Camacho-Cooper_2022,Du_2024}, and collective emission~\cite{Yu_2017}.

Moiré heterostructures have also expanded our toolbox for quantum simulation, both for realizing electronic phases and for studying excitons in the quantum degenerate regime. In these systems, the interplay between spin and valley degrees of freedom—central to spintronics and valleytronics—offers a powerful platform for engineering mixed quantum systems. The band structure can be tuned to host electrons and holes with distinct spin and valley flavors~\cite{Xu_2014, Schaibley_2016,Jin_2019_nov}. This tunability allows for the selective trapping of fermions and bosons in different moiré minibands or layers, enabling the realization of exotic quantum mixtures. For instance, spin-polarized excitons (bosons) can coexist with itinerant electrons (fermions), forming Bose-Fermi mixtures \cite{Bastarrachea_2021M,Bastarrachea_2021M1,Julku_2021}, while valley-selective occupation of fermionic states in different layers can give rise to Fermi-Fermi mixtures\cite{Burg_2019,Liu_2020,Wang_2020,Ghiotto_2021}.

Similarly, excitons with differing configurations behave as distinguishable bosonic species, enabling the creation of Bose-Bose mixtures. In bilayer systems, excitons can form either from the binding of an electron and a hole within the same layer (intralayer excitons) or from carriers residing in opposite layers (interlayer excitons)~\cite{Rivera_2015,Torun_2018,Alexeev_2019,Ruiz-Tijerina_2019,Tartakovskii_2020,Shimazaki_2020,Jiang_2021}. The direct or indirect character of these excitons gives rise to contrasting properties~\cite{Huang_2022,Miller_2017_a,Zhang_2023}. Typically, intralayer excitons possess strong oscillator strength but exhibit weak exciton-exciton interactions, whereas interlayer excitons couple weakly to light~\cite{Wang_2018_a,Regan_2022_a}. However, their spatially indirect nature endows them with a permanent dipole moment that strongly enhances exciton-exciton interactions. Bose-Bose mixtures offer a promising platform for realizing quantum phases such as supersolids~\cite{Trousselet_2014,Julku_2022}, quantum droplets~\cite{Petrov_2015}, fluctuation-governed gases~\cite{Boudjem_2018}, and bound states of photons~\cite{Firstenberg_2013,Carusotto_2013,Takemura_2014}.
\begin{figure}[h]
\centering
\includegraphics[width=.6\columnwidth]{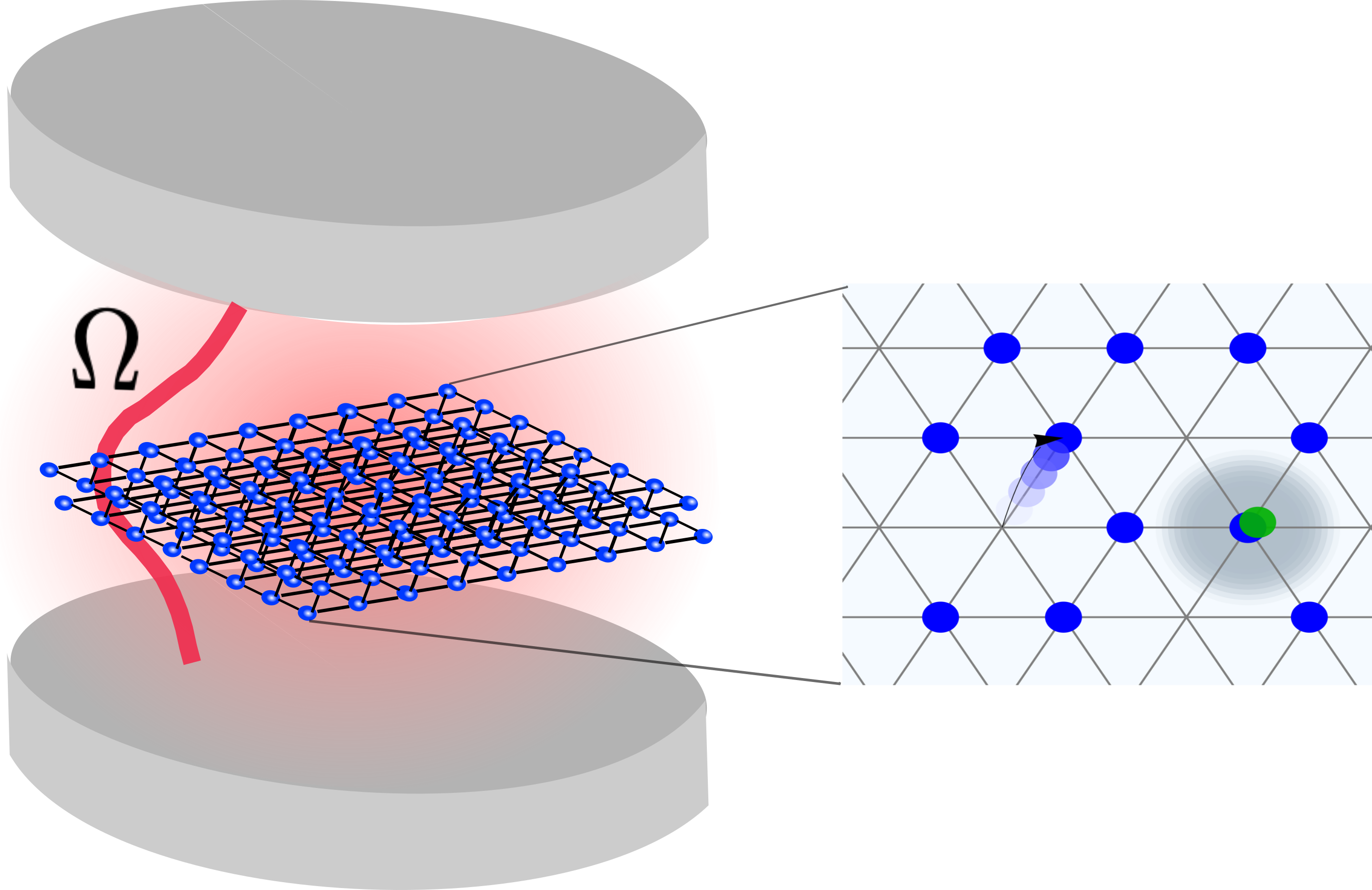}
\caption{We consider a semiconductor bilayer hosting intralayer and interlayer excitons confined to a moiré superlattice. Interlayer excitons form a Bose-Einstein condensate of hard-core bosons and do not couple to light. A weak light probe injects photons into the cavity forming intralayer exciton polaritons. }
\label{Model}
\end{figure}

Motivated by recent progress in the field, we study a highly population-imbalanced Bose-Bose mixture, consisting of a Bose-Einstein condensate (BEC) of dark-state interlayer excitons confined to a moiré superlattice, and coupled to a minority population of intralayer exciton-polaritons. We develop a variational approach to investigate the formation of polaritons in the presence of strong exciton-exciton interactions and find that, along with an energy shift of the avoided crossing, a polariton branch emerges as a consequence of a repulsive bound state between an interlayer and an intralayer exciton. The optical response of the system is highly sensitive to the filling factor of the BEC of dark-state excitons, which drives the system from a non-interacting to a weakly interacting, and eventually to a strongly interacting regime.

The study of moiré polaritons coupled to a BEC of interlayer excitons unveils new mechanisms to probe and detect dark-state condensation, and reveals the 
emergence of a repulsive polariton branch that is highly sensitive to the filling factor of the dark BEC.


{\it System.-}\label{sec:model} 
Consider a van der Waals semiconductor hetero-bilayer hosting both intralayer and interlayer excitons. A highly population-imbalanced Bose-Bose mixture is formed by a majority Bose-Einstein condensate of ground-state interlayer excitons, coupled to a minority population of intralayer exciton-polaritons. A periodic moir\'e potential emerges due to the slight lattice mismatch of the heterostructure, confining both exciton species to the sites of a moir\'e superlattice. The hetero-bilayer is then embedded into a microcavity, as illustrated in Fig.~\ref{Model}.

In general, intra- and interlayer moir\'e excitons may localise at different sites of the moir\'e supercell~\cite{Yu_2017,multifaceted,Naik2022}. Here, we assume that both species localise at the same position, and describe them using the Bose-Bose Hubbard model
\begin{equation}
    \hat{H}= \hat{H}_{X}+\hat{H}_{X-Y}+\hat{H}_{P}.
\end{equation}
The interlayer exciton terms of the Hamiltonian are
\begin{gather} \label{X_exciton_Ham}
     \hat{H}_{X}=-t_X\sum_{\langle \mathbf{r},\mathbf{r}^{'}\rangle}\big(\hat{X}^{\dagger}_{\mathbf {r}}\hat{X}_{\mathbf{r}{{'}}}+\mathrm{h.c}\big)-\mu_{B}\sum_{\mathbf r}\hat{X}^{\dagger}_{\mathbf r}\hat{X}_{\mathbf r} \\  \nonumber
    +\frac{U_{XX}}{2}\sum_{\mathbf r}\hat{X}^{\dagger}_{\mathbf r}\hat{X}^{\dagger}_{\mathbf r}\hat{X}_{\mathbf r}\hat{X}_{\mathbf r}, 
\end{gather}
where $\Xhat_{\rr}^\dagger$ creates an interlayer exciton at superlattice site $\rr$, $t_X$ is the tunnelling coefficient between nearest-neighbor (NN) sites, and $\mu_B$ is the chemical potential. The exciton-exciton interaction is assumed to be local and characterised by $U_{XX}.$ Due to their spatially indirect nature, interlayer excitons possess a permanent dipole moment, enhancing $U_{XX}$ and making it repulsive. Henceforth, we take $U_{XX}$ as the largest relevant energy scale in the problem. This forbids the occupancy of a moir\'e site by more than one interlayer exciton in the ground state, and justifies treating them as hard-core bosons.

Intralayer exciton-polaritons, formed by the hybridisation of cavity photons and intralayer excitons, are described by
\begin{gather}
        \hat{H}_P=-t_Y\sum_{\langle \mathbf r,\mathbf r^{'}\rangle}\big(\hat{Y}^{\dagger}_{\mathbf r}\hat{Y}_{\mathbf r{{'}}}+\mathrm{h.c}\big)+\sum_{\mathbf k}\omega_{\mathbf k}\hat{a}^{\dagger}_{\mathbf k}\hat{a}_{\mathbf k}+\\ \nonumber
  +\Omega\sum_{\mathbf k}(\hat{Y}_{\mathbf k}^{\dagger}\hat{a}_{\mathbf k}+\hat{a}^{\dagger}_{\mathbf k}\hat{Y}_{\mathbf k}),
  \label{HX}
\end{gather}
where $\Yhat^\dagger_{\rr}$ is the intralayer exciton creation operator, and $t_Y$ is the NN tunnelling coefficient for intralayer excitons. $\ahat_{\kk}^\dagger$ creates a cavity photon with momentum $\kk$ and energy $\omega_{\kk}$, which hybridises with intralayer excitons in the rotating-wave approximation with Rabi frequency $\Omega$. Note that we have introduced the intralayer-exciton moir\'e-Bloch operators $\Yhat_{\kk}=\sum_{\rr}\tfrac{e^{i\kk\cdot\rr}}{\sqrt{N}}\Yhat_{\rr}$, with $N$ the number of moir\'e supercells in the sample. No exciton-photon interaction is considered for interlayer excitons, since their oscillator strength is known to be negligible due to their spatially-indirect nature~\cite{PhysRevLett.115.187002,Ross2017,PhysRevB.97.035306}.

Finally, we consider a repulsive on-site interaction between intra- and interlayer excitons
\begin{gather}\label{eq:UXY}
    \hat{H}_{X-Y}= U_{XY}\sum_{\mathbf r}\hat{X}^{\dagger}_{\mathbf r}\hat{X}_{\mathbf r}\hat{Y}^{\dagger}_{\mathbf r}\hat{Y}_{\mathbf r}.
\end{gather}
In real 2D heterostructures, hybridisation between intra- and interlayer excitons can occur due to interlayer electron- or hole tunnelling, in cases where the two exciton energies are sufficiently close~\cite{Alexeev_2019,PhysRevLett.132.076902}. In the following we assume a large detuning between intra- and interlayer excitons, such that their hybridisation can be neglected, and the two species interact only through the repulsive term \eqref{eq:UXY}.

{\it Methods.-} Our approach is based on two main considerations. First, we assume a Bose-Einstein condensate (BEC) of hard-core, dark-state interlayer excitons, and compute their mean-field solution and collective excitations. This allows us to reexpress the inter-species interaction \eqref{eq:UXY} in terms of the collective excitations of the condensate. 
Second, we deploy a variational, two-body \emph{ansatz} to describe non-perturbatively the ground-state. This procedure, will allow us to understand the formation of moiré polaritons coupled to a bath of condensed dark excitons.

We follow closely the formalism introduced in Refs.\ \cite{Analytical_2002,Santiago-García_2023,Santiago-García_2024}, slightly modified for a triangular moir\'e superlattice, typical of transition metal dichalcogenide (TMD) hetero-bilayers. We enforce the hard-core boson constrain by mapping the exciton operators' action on the restricted occupancy subspace onto that of spin-$\nicefrac{1}{2}$ operators: $\hat X^\dagger_{\mathbf r}\rightarrow \hat S^+_{\mathbf r}$ and $\hat X_{\mathbf r}\rightarrow \hat S_{\mathbf r}^-$, with $\hat S^\pm_{\mathbf r}$ the spin-$\nicefrac{1}{2}$ ladder operators at site $\rr$. Null occupancy of moir\'e site $\rr$ by interlayer excitons is represented by state $\ket{\downarrow}_{\rr}$, whereas single occupancy is $\ket{\uparrow}_{\rr}=S_{\rr}^+\ket{\downarrow}_{\rr}$. Then, the interlayer exciton BEC at the mean-field level is
\begin{gather}
    |\psi(\theta)\rangle= \prod_{\mathbf r}\left(\sin\left(\frac{\theta}{2}\right)+\cos\left(\frac{\theta}{2}\right)\hat{S}_{\mathbf r}^{+}\right)|0\rangle,
\end{gather} where $|0\rangle= \prod_{\mathbf r}|\downarrow \rangle_{\mathbf r}$
represents the vacuum state, 
and $\theta$ can be related to the chemical potential and filling factor straightforwardly, by minimizing the energy following a variational approach,  $\delta \langle \psi |\hat{H}_{X}| \psi \rangle/ \delta\theta =0$. We find that $\cos\theta = \mu_{B}/6t_{X}$, which defines the filling factor $n=(1+\cos{\theta})/2$. 

Next, we obtain the collective excitations following a similar approach as Refs.~\cite{Analytical_2002,Santiago-García_2023,Santiago-García_2024}. In terms of the excitation operators $\gammahat_{\kk}^\dagger$ and $\gammahat_{\kk}$, the interlayer exciton Hamiltonian of Eq.~\eqref{X_exciton_Ham} takes the simple form
\begin{equation*}
\hat{H}_{X}=\sum_{\mathbf k}E(\mathbf k)\hat{\gamma}_{\mathbf k}^{\dagger}\hat{\gamma}_{\mathbf k},
\end{equation*} 
with $E(\mathbf k)= \sqrt{\alpha_{\mathbf k}^{2}-\beta_{\mathbf k}^{2}}$ the energy of the collective mode of crystal momentum $\kk$. We have defined
\begin{equation*}
\alpha_{\mathbf k}= \frac{\epsilon_{\mathbf k}^{X}}{4}\big(1+\cos^{2}\theta\big)+3t_{X},\, \beta_{\mathbf k}=-\frac{\epsilon_{\mathbf k}^{X}}{4} \sin^{2}\theta,    
\end{equation*}
with the bare exciton energy $\epsilon_{\mathbf k}^{X}=-2t_{X}[\cos(ak_{x})+2\cos(\tfrac{ak_{x}}{2})\cos(\tfrac{\sqrt{3}}{2}ak_{y})]$, and $a$ the moir\'e superlattice constant. 

Under the same transformations for the majority hard-core bosons, the interaction term \eqref{eq:UXY} takes the form

 \begin{widetext}
 \begin{equation}
 \begin{split}
 \label{Hcompleto}
     \hat{H}_{X-Y}=&\epsilon_{MF}\sum_{\mathbf k}\hat{Y}_{\mathbf k}^{\dagger}\hat{Y}_{\mathbf k} -\frac{U_{XY}}{2\sqrt{N}}\sin\theta \sum_{\mathbf k,\mathbf q}\big(u_{\mathbf q}-v_{\mathbf q}\big)\big(\hat{\gamma}_{-\mathbf q}^{\dagger}+\hat{\gamma}_{\mathbf q}\big)\hat{Y}_{(\kk+\qq)}^{\dagger}\hat{Y}_{\mathbf k} \\
    &-\frac{U_{XY}}{N}\cos\theta\sum_{\mathbf k,\mathbf k^{'},\mathbf q}\Big[u_{(\mathbf k^{'}+\mathbf q)}u_{\mathbf k{{'}}}\hat{\gamma}_{(\mathbf k^{'}+\mathbf q)}^{\dagger}\hat{\gamma}_{\mathbf k^{'}}-u_{(\kk'+\qq)}v_{\mathbf k^{'}}\hat{\gamma}_{\mathbf (\kk^{'}+\mathbf q)}^{\dagger}\hat{\gamma}_{-\mathbf k^{'}}^{\dagger} \\
     &\qquad\qquad\qquad\qquad-v_{\mathbf (\kk^{'}+\qq)}u_{\mathbf k^{'}}\hat{\gamma}_{-(\mathbf k^{'}+\mathbf q)}\hat{\gamma}_{\mathbf k^{'}}+v_{(\mathbf k^{'}+\mathbf q)}v_{\mathbf k^{'}}\hat{\gamma}_{-(\mathbf k^{'}+\mathbf q)}\hat{\gamma}_{-\mathbf k^{'}}^{\dagger}\Big]\hat{Y}_{(\mathbf k-\mathbf q)}^{\dagger}\hat{Y}_{\mathbf k},
 \end{split}
 \end{equation}
 \end{widetext}
with $\epsilon_{MF}=U_{XY}(1+\cos{\theta})/2$ is the mean-field energy shift of the $Y$ exciton due to the coupling to the $X$ excitons. Finally $u_{\mathbf k}$ and $v_{\mathbf k}$ are the coherent factors
\begin{equation*}
    u_{\mathbf k}=\sqrt{\frac{1}{2}+\frac{\alpha_{\mathbf k}}{2\sqrt{\alpha_{\mathbf k}^{2}-\beta_{\mathbf k}^{2}}}},\,
    v_{\mathbf k}=\sqrt{-\frac{1}{2}+\frac{\alpha_{\mathbf k}}{2\sqrt{\alpha_{\mathbf k}^{2}-\beta_{\mathbf k}^{2}}}},
\end{equation*}
of the Bogoliubov transformations relating the collective excitations to the hard-core boson operators. It is important to stress that the Hamiltonian in Eq.~\eqref{Hcompleto} contains the mean-field interaction, the Fr\"ohlich-type interaction, and the beyond-Fr\"ohlich terms necessary to account for the binding between $X$ and $Y$ atoms. The latter are responsible for the formation of repulsive bound state.


Finally, we introduce a variational \emph{ansatz} to describe a single optical excitation coupled to the dark exciton BEC:
  \cite{Sidler2017} 
\begin{align} \nonumber 
    &|P_{\mathbf 0}\rangle=\\  &\bigg(\phi_{\mathbf 0}^{a}\hat{a}_{\mathbf 0}^{\dagger}+\varphi_{\mathbf 0}^{y}\hat{Y}_{\mathbf 0}^{\dagger}+\sum_{\mathbf p}\varphi_{\mathbf p}^{y}\hat{Y}_{\mathbf p}^{\dagger}\hat{\gamma}_{-\mathbf p}^{\dagger}+\sum_{\mathbf p}\phi_{\mathbf p}^{a}\hat{a}_{-\mathbf p}^{\dagger}\hat{\gamma}_{\mathbf p}^{\dagger}\bigg)|\psi(\theta)\rangle.
\end{align}
which considers the superposition of a cavity photon, and a $Y$ intralayer exciton dressed by the collective excitations of the $X$ interlayer excitons. The variational parameters $\phi_{\mathbf 0}^{a}$, $\varphi_{\mathbf 0}^{y}$, $\varphi_{\mathbf p}^{y}$, and $\phi_{\mathbf p}^{a}$ are determined by minimizing the energy $\delta \langle P_{\mathbf 0}|\hat{H}-E|P_{\mathbf 0}\rangle/\delta x=0$, where $x={\phi_{\mathbf 0}^{a*}, \varphi_{\mathbf 0}^{y*}, \varphi_{\mathbf p}^{y*}, \phi_{\mathbf p}^{a*}}$. This approach is inspired by previous studies of Fermi polaron-polaritons~\cite{Sidler2017} in conventional two-dimensional semiconductors, now adapted to moiré polaritons coupled to a dark BEC of hard-core bosons. We assume that the excitons couple only to the $\kk=\boldsymbol{0}$ mode of the cavity.  The effective mass of the cavity photons is orders of magnitude smaller than of the excitons, thus defining an ultra-small light cone for the exciton-photon interaction. This justifies the assumption that only $\mathbf k=0$ excitons couple to the cavity mode.


The resulting set of variational equations is
\begin{align}
    \omega_{\mathbf 0}\phi_{\mathbf 0}^{a}+\Omega \varphi_{\mathbf 0}^{y}&= E\phi_{\mathbf 0}^{a} \label{photon_zero} \\
    \Omega \phi_{\mathbf 0}^{a}+\big[\epsilon_{MF}+\epsilon_{\mathbf 0}^{y}-\frac{U_{XY}}{N}\cos\theta\sum_{\mathbf p}v_{\mathbf p}^{2}\big]\varphi_{\mathbf 0}^{y}&\nonumber  \\
    -\frac{U_{XY}}{2\sqrt{N}}\sin\theta\sum_{\mathbf p}\varphi_{\mathbf p}^{y}\big(u_{\mathbf p}-v_{\mathbf p}\big)&=E\varphi_{\mathbf 0}^{y} \label{exciton_zero} \\
    -\frac{U_{XY}}{2\sqrt{N}}\sin\theta\big(u_{\mathbf p}-v_{\mathbf p}\big)\varphi_{\mathbf 0}^{y}+\big(E(\mathbf p)+\epsilon_{\mathbf p}^{y}&+\epsilon_{MF} \nonumber\\
    -\frac{U_{XY}}{N}\cos\theta\sum_{p}v_{\mathbf p}^{2}\big)\varphi_{\mathbf p}^{y}-\frac{U_{XY}}{N}&\cos\theta\nonumber \\
    \sum_{\mathbf p^{'}}\big(u_{\mathbf p}\sum_{\mathbf p^{'}}u_{\mathbf p^{'}}\varphi_{\mathbf p^{'}}^{y}+v_{\mathbf p}\sum_{\mathbf p^{'}}v_{\mathbf p^{'}}\varphi_{\mathbf p^{'}}^{y}\big)&= E\varphi_{\mathbf p}^{y} ,\label{exciton_k}
\end{align}

\begin{figure*}[t]
 {\includegraphics[width=1.8\columnwidth]{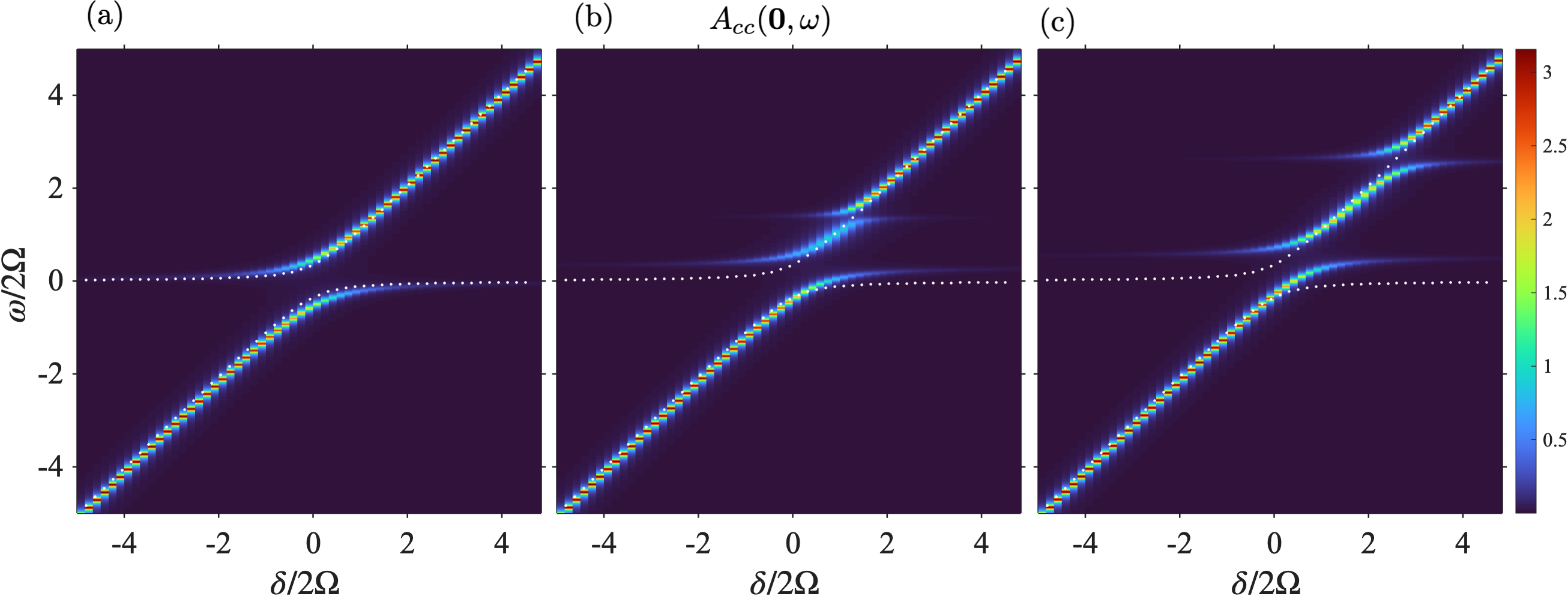}}
\caption{Spectral function as a function of the parameter $\delta$ for different densities of the $X$ excitons and different ratios of $U_{XY}/\Omega$.
For $\theta = \pi/2$ (half-filling): a) $U_{XY}/\Omega = 0$, b) $U_{XY}/\Omega = 2$, and c) $U_{XY}/\Omega = 5$. We take $t = 0.1\Omega$. The on-site energy of the $Y$ exciton is fixed to zero, and shifting its value would only result in a rigid shift of the entire spectrum.
 }
\label{Spectral_function_det}
\end{figure*}

where $\epsilon_{\mathbf k}^{Y}=-2t[\cos(k_{x}a)+2\cos(\frac{k_{x}a}{2})\cos(\frac{\sqrt{3}}{2}k_{y}a)]+6t_{Y}$ is the energy of the bare intralayer excitons. We numerically solve this set of equations. 

To study the properties of the system we introduce the spectral function of the cavity photons defined as
\begin{equation}\label{eq:spectralfunc}
    A_{cc}(\mathbf k=0,\omega)=-\frac{1}{\pi}\mathbf{Im} \bigg[ \sum_{n=0}\frac{|\langle \phi_{\mathbf 0}^{a}|\Psi_n\rangle|^{2}}{\omega - E_{n}+i \eta} \bigg],
\end{equation}
 with $|\phi_{\mathbf 0}^{a}\rangle=\hat a^\dagger_{\mathbf 0}|\psi(\theta)\rangle,$ a bare photon on top of the BEC. Here, $\{|\Psi_n\rangle\}$ is the set of eigenstates with eigenenergies $E_n$ obtained from solving the set of equations~\eqref{photon_zero}-\eqref{exciton_k}. We add a finite broadening $\eta$ to visualize the spectral function. Since we focus on the zero-momentum photon mode, we henceforth abbreviate the spectral function \eqref{eq:spectralfunc} as $A(\omega)$.

{\it Results.-}\label{sec:LME} Now, we study the system as function of its tunable parameters. First, we study the spectral function of the cavity photons as a function of the detuning $\delta = \omega_0 - \epsilon_{\mathbf 0}^Y$ between the photon and intralayer exciton energy minima, considering several values of both the interspecies interaction $U_{XY}$ and the filling factor $n$. For our numerical calculations we take $t_X=t_Y=t.$

Figure \ref{Spectral_function_det} shows the spectral function of cavity photons at half-filling ($n=1/2$ and $\theta=\pi/2$). Figure~\ref{Spectral_function_det}(a) corresponds to the non-interacting ($U_{XY}=0$) case, where we recover the standard polariton picture: two  branches---the lower (LP) and upper (UP) polaritons---with energies $\omega_{UP/LP} = \frac{\delta \pm \sqrt{\delta^2 + 4\Omega^2}}{2}$. For reference, these branches are shown with white dotted lines in all panels of Fig.\ \ref{Spectral_function_det}.

For $U_{XY}/\Omega=2$ [Fig.~\ref{Spectral_function_det}(b)], we observe significant deviations from the conventional polariton spectrum. First, the avoided crossing between the main polariton branches shifts toward positive detunings. This can be understood even at the mean-field level, where the energy of the bright, intralayer exciton shifts to $\epsilon_\mathbf{0} + nU_{XY}$ due to its coupling to the medium. Second, the polariton branches split, and a narrow and blurred intermediate polariton state emerges.

For strong interactions [Fig.~\ref{Spectral_function_det}(c)], $U_{XY}/\Omega=5$, we obtain an even more dramatic departure from the conventional polaritons. Not only does the main avoided crossing shift to larger positive values, but the third intermediate polariton state also becomes well-defined. The emergence of this middle polariton branch is a non-perturbative result, and its physical origin is an underlying two-body bound state between an intra- and an interlayer exciton. Interestingly, the energy of this bound state lies above the energy of two free excitons; that is, it corresponds to a repulsive bound state. To capture the physics of the repulsive bound state, it is not only necessary to include the beyond-Fr\"ohlich terms in Eq.~\eqref{Hcompleto}, but to employ a non-perturbative approach such as our variational ansatz~\cite{Ding2024}.

\begin{figure}[b]
 {\includegraphics[width=.8\columnwidth]{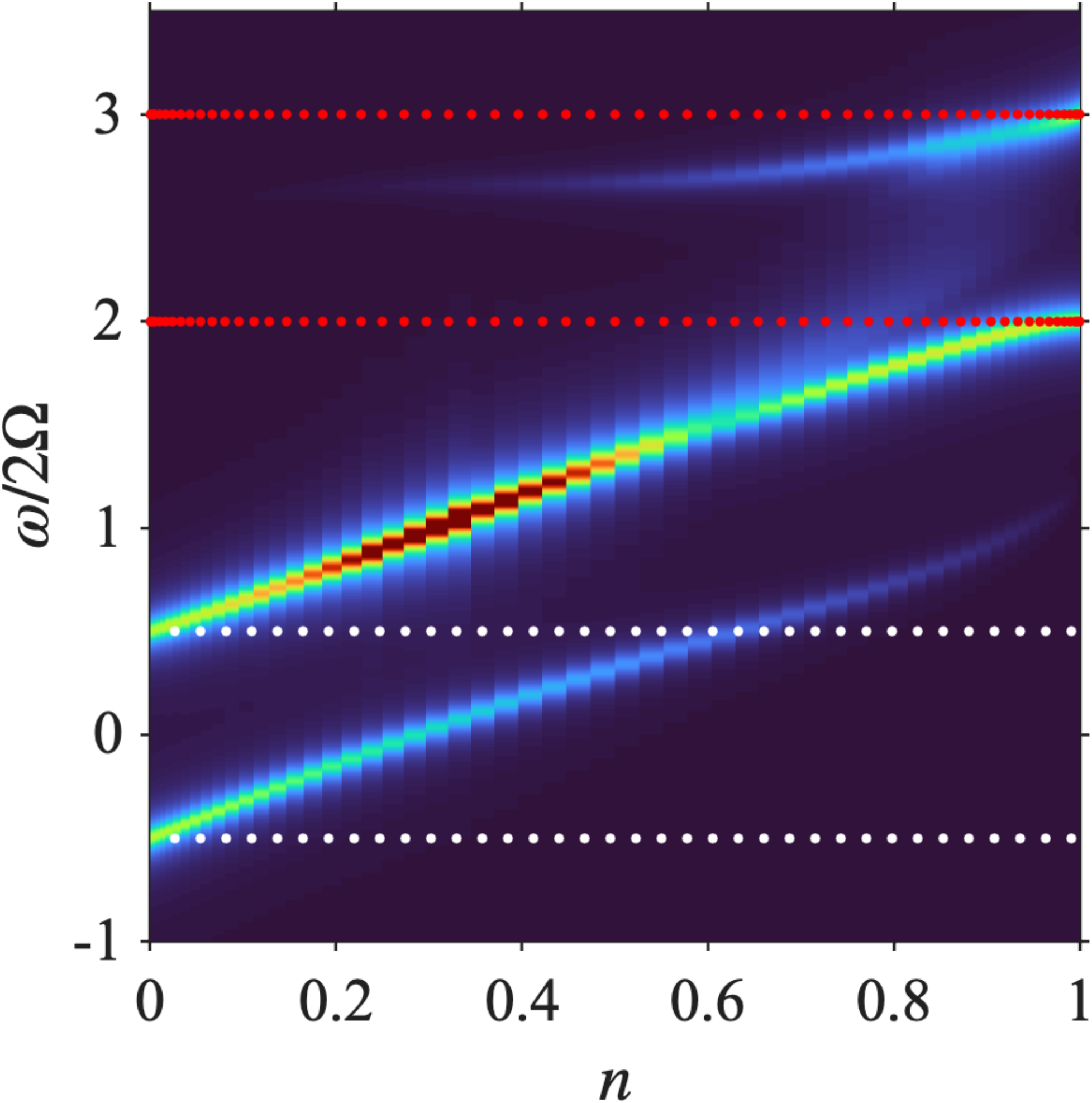}}
\caption{Spectral function as function of the parameter filling factor of the dark BEC. We take $U_{XY}/\Omega=5$, $\delta=nU_{XY}$, and $t=.1\Omega$.}
\label{Spectral_function_theta}
\end{figure}

To understand the role of dark BEC density, we show the spectral function of the cavity photons as a function of the filling factor. As shown in Fig.~\ref{Spectral_function_det}, interactions shift the resonance condition—that is, the avoided crossing is shifted to positive detunings. Thus, in Fig.~\ref{Spectral_function_theta}, we take $\delta = nU_{XY}$ to ensure that excitons and photons remain maximally coupled.

Figure~\ref{Spectral_function_theta} shows that the system evolves from a non-interacting to a weakly interacting regime as a function of the filling factor. For $n \ll 1$, we recover the two non-interacting lower and upper polaritons. As the filling factor is slightly increased, we observe that both polariton states shift to positive energies, and the lower polariton cedes spectral weight to the upper polariton. 

Around $n \sim 1/3$, not only is spectral weight transferred to the upper branch, but a new polariton state also emerges at even higher energies. For $n \sim 0.75$, the lower polariton completely loses its spectral weight, and the two upper polariton branches dominate. 

Finally, near $n \sim 1$, the system again exhibits only two polariton branches, with energies $\omega = \omega_{\text{UP}/\text{LP}} + nU_{XY}$, marked by the red dots. That is, in this regime, the energy cost of adding a $Y$ exciton to the system is $U_{XY}$, the energy of double occupation: a $X$ and a $Y$ exciton. 

Interestingly, the rich interplay between the various polariton states occurs predominantly near half-filling. Note that $ n \ll 1 $ and $ n \sim 1 $ can be understood as lattices filled predominantly with holes and particles, respectively. The competition between the polariton branches arises from the mobility of the bosons forming the dark-state BEC, which occurs predominantly near $n \sim \frac{1}{2} $.

In the context of ultracold gases, stable repulsive two-body bound states have already been experimentally demonstrated, where confinement in optical lattices plays a key role ~\cite{winkler2006repulsively}. In monolayer semiconductors, the splitting of polariton branches has only been reported in the context of attractive biexciton and trion states~\cite{Sidler2017,Takemura_2014,Navadeh2024,Takemura2017,Bastarrachea2019,Ding2024,Bastarrachea2024}. Conventional excitons are prevented from forming repulsive two-body bound states due to the free 2D dispersion. In stark contrast, moiré excitons can be tightly confined to a moiré superlattice, and in the presence of strong interactions, a repulsive two-body bound state can emerge. One of the main results of this study is the prediction of a polariton branch arising from this repulsive bound state.

{\it Experimental implementation.-} The Bose-Bose mixture discussed in this letter naturally arises in TMD hetero-bilayers with type II (staggered) band alignment, where interlayer excitons typically appear at lower energies than the intralayer excitons of the constituting monolayers, by several hundreds of meV~\cite{Zhang_2017,Rivera2018,Liu_2020}. This detuning is sufficient to suppress hybridisation between the two species, and to promote a large population imbalance favouring interlayer excitons, especially at low temperatures. BEC of interlayer excitons in these heterostructures has been recently observed~\cite{wang2019evidence}, and is made possible by the long lifetime of this exciton species, promoted by their weak interaction with light~\cite{PhysRevLett.115.187002,Ross2017,PhysRevB.97.035306}, inefficient intervalley scattering ~\cite{RiveraScience2016}, and slow recombination rate by nonradiative processes~\cite{Vialla_2019,Cai2024}. The interaction $U_{XX}$ in TMD-based hetero-bilayers has been measured at~\cite{Park_2023} $30-40\,{\rm meV}$. On the other hand, the Rabi frequency for intralayer exciton-polaritons has been measured at~\cite{Dufferwiel_2015,flatten2016room,sasha2023} $20-70\,{\rm meV}$, whereas for  moiré polaritons it has been reported as $\Omega\sim 5-10\text{meV}$~\cite{Zhang_2021}. Although to our knowledge $U_{XY}$ has not been measured experimentally, a conservative estimate of $U_{XY}\sim 5\, {\rm meV},$ since in these systems $t\ll U_{Y},$ excitons remain in the strongly interacting regime.  Thus, the repulsive bound state is expected to appear in the optical spectrum as a fine but sharp feature of the upper polariton branch, at energies above that of the main avoided crossing.

An ideal candidate is the parallel-stacked MoSe${}_2$/WSe${}_2$ hetero-bilayer, where \emph{ab initio} calculations predict\cite{Yu_2017} that the interlayer excitons and the WSe${}_2$ intralayer excitons will localise at AB staking regions of the moir\'e superlattice.  Also of interest is parallel-stacked MoS${}_2$/WSe${}_2$, where the interlayer exciton only localises at AB stacking regions, like the WSe${}_2$ intralayer exciton, under the influence of a critical out-of-plane electric field. Tuning this field allows some measure of control over the width of the confined interlayer exciton wave function, while leaving the intralayer exciton state unaffected, thus modifying $U_{XY}$ while keeping the intralayer exciton parameters constant.


{\it Conclusions.-} Moiré superlattices in van der Waals heterostructures offer a versatile and highly tunable platform for realizing quantum Bose-Bose mixtures. These systems provide a promising platform for exploring emergent many-body phenomena and for engineering quantum simulators capable of emulating complex systems relevant to condensed matter, atomic physics, and high energy physics, providing unprecedented access to regimes typically unreachable in conventional systems.

In this work, we have explored the interplay between moiré polaritons coupled to a Bose-Einstein condensate of dark excitons confined in a moiré superlattice. By developing a variational approach, we have demonstrated that the presence of strong exciton-exciton interactions dramatically alters the optical response of the system. Notably, the emergence of a repulsive bound state gives rise to a new polariton branch, an additional avoided crossing, and to the formation of repulsive moiré polaron-polaritons. Our findings show that the spectral response of the cavity photons is highly sensitive to the filling factor of the underlying condensate of dark excitons.  
 
 Our results highlight the ability to use moiré polaritons to probe dark-states of matter which are typically hard to access. The sensitive of moiré polaritons to the underlying dark condensate unveils a mechanism to characterize quantum gases opening also new directions for engineering light-matter systems with tunable interactions and for the realization of exotic polaritons in strongly correlated moiré heterostructures.



\acknowledgments
 M.S-G. acknowledges SECIHTI, Mexico, for his scholarship.  A.C-G. acknowledges financial support from UNAM DGAPA PAPIIT Grant No. IA101325, Project CONAHCYT No. CBF2023-2024-1765 and PIIF25.
 S.G.C-L. acknowledges financial support from UNAM DGAPA PAPIIT Grant No.
TA100724. D.A.R-T.\ acknowledges funding from UNAM DGAPA PAPIIT grant No.\ IN114125.
 
\bibliography{bib}

\end{document}